\documentclass{jpsj3}
\title{The high-lying  $^6$Li levels at excitation energy around 21 MeV}

\author{
Orest \textsc{Povoroznyk}\thanks{E-mail address:  orestpov@kinr.kiev.ua}, Olga K. \textsc{Gorpinich},  Olexiy O. \textsc{Jachmenjov}, Hanna V.\textsc{Mokhnach}, Oleg \textsc{Ponkratenko},
 Giuseppe \textsc{Mandaglio}$^{1,2}$,
 Francesca \textsc{Curciarello} $^{1,2}$, Veronica \textsc{De Leo}$^{1,2}$,  Giovanni \textsc{Fazio}$^{1,2}$, 
Giorgio \textsc{Giardina}$^{1,2}$\thanks{E-mail address: giardina@nucleo.unime.it} 
 }

\inst{Institute for Nuclear Research, National Academy of Science of Ukraine, 03680 Kiev, Ukraine\\
$^1$Dipartimento di Fisica dell'Universit\`a  di Messina, 98166 Messina, Italy\\$^2$ Istituto Nazionale di Fisica Nucleare, Sezione di Catania, 95123 Catania, Italy}

\abst{The $^3$H+$^3$He cluster structure in $^6$Li was investigated by the $^3$H($\alpha$,$^3$H $^3$He)n kinematically complete experiment at the incident energy  $E_\alpha$ = 67.2 MeV. We have observed two resonances at $E_x^*$ = 21.30 and 21.90 MeV which are consistent with the $^3$He($^3$H,~$\gamma$)$^6$Li analysis in the Ajzenberg-Selove compilation. Our data are compared with the previous experimental data and the RGM and CSRGM calculations.}

\kword{$^4$He-induced reaction, $^3$H-target, high-lying $^6$Li levels, $E^*$ and $\Gamma$ spectroscopic parameters, 3+3 cluster decay}

\begin{document}
\maketitle

\section{Introduction} 

The study of  three-body reactions, using the inclusive or bidimensional spectra, is a suitable way of obtaining information on the reaction dynamics and nuclear spectroscopy. The inclusive  spectra frequently show a strong and continuous background from other states of the nucleus of interest, the decay of other nuclei produced in the competing reaction channels, and the statistical three-body break-up. The kinematically complete measurement provides a powerful way to study three-body resonances since these backgrounds  are largely reduced.

The spectroscopic information on low-lying states in light nuclei is known with sufficient accuracy, while that on high-lying states is less known. Generally, these are unbound with short-life values, and thus their energy widths are significantly large. Therefore, it would be difficult to obtain their excitation energies with high accuracy.

 The $^6$He and $^6$Li nuclei are of remarkable theoretical and experimental interest because they have cluster structures of alpha+2n and alpha+d, respectively. Particularly, for the $^6$He nucleus the core is an alpha  and the halo consists of two neutrons ($\alpha$+2n structure), while for the $^6$Li nucleus the core is the same  and the halo is a deuteron ($\alpha$+d structure). At higher excitation energies, other structures, like $^3$H+$^3$H (t + t) for  $^6$He and $^3$He+$^3$H  ($\tau$ + t) for $^6$Li, have to be considered around their threshold energies of 12.203 and 15.795 MeV, respectively. 
Thus it is very interesting to investigate the high-lying $^6$He  and $^6$Li  cluster states in order to obtain their spectroscopic information.

\begin{figure}[h]
\begin{center}
\vspace{-1.7cm}
\resizebox{0.95\textwidth}{!}{\includegraphics{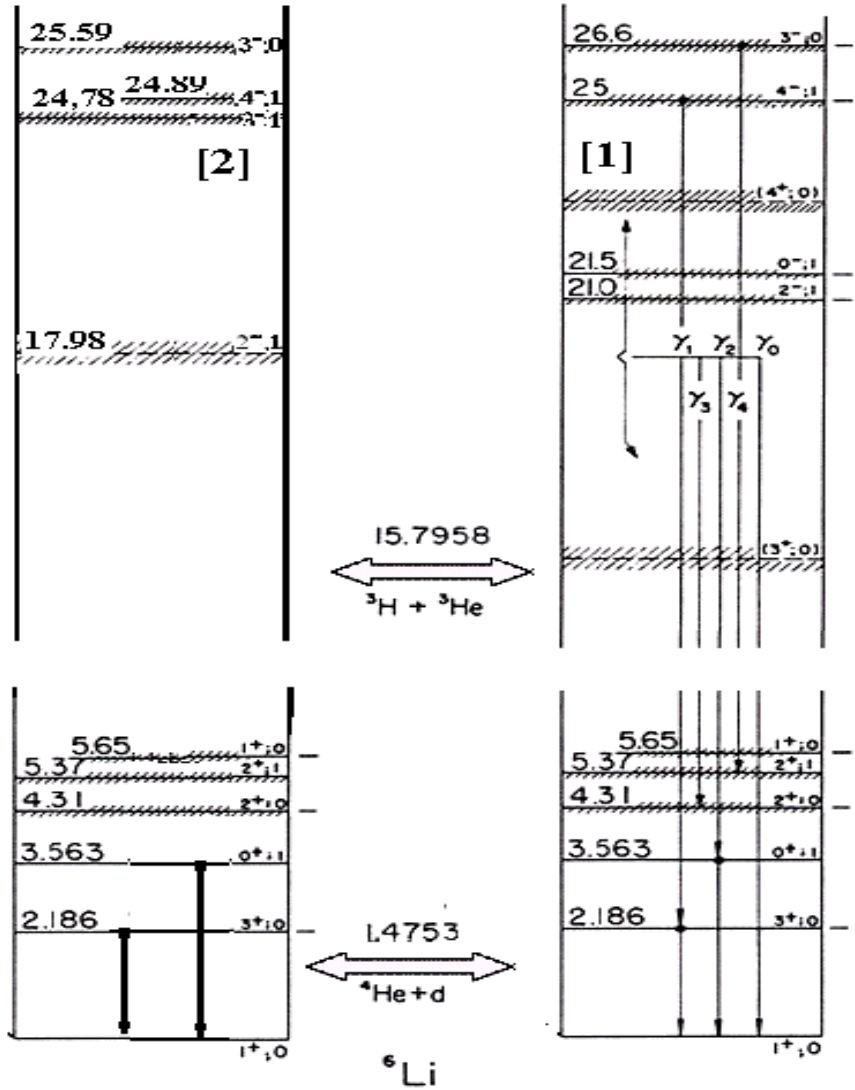}}
\end{center}
\vspace{-5.0 cm}
\caption{Energy levels of $^6$Li as compiled by Ajzenberg-Selove \cite{Ajzenberg1984} and Tilley {\it et al.} \cite{Tilley02}.}
\label{fig1}
\end{figure}

   Fig. \ref{fig1} represents the $^6$Li energy levels given by Ajzenberg-Selove\cite{Ajzenberg1984} and Tilley {\it et al.}\cite{Tilley02}. A significant discrepancy is found at excitation energies higher than the t+$\tau$ decay threshold of $E_x$ = 15.795 MeV. It should be noted that short-life  $^6$Li levels have been observed at $E_x$ = 18 and 22 MeV with $\Gamma$~=~5.0 $\pm$ 0.5 and 8 $\pm$ 1 MeV, respectively, in the recent $^7$Li($^3$He, $\alpha$)$^6$Li  study\cite{Nakayama} at E($^3$He)=450 MeV.

Thompson and Tang\cite{Thompson} theoretically studied the existence of trinucleon ($\tau$ and t) clusters in $^6$Li in the framework of the LS coupling analysis with the RGM calculations\cite{Thompson}. They have predicted a P doublet ($^1$P$_1$ and $^3$P$_{0,1,2}$) and a F doublet ($^1$F$_3$ and $^3$F$_{2,3,4}$) at $E_{\rm x}^*$ $\simeq$ 22 and 29 MeV, respectively. On the contrary, Vlastou {\it et al.} \cite{Vlastou} have reported $^3$P$_2$ , $^3$P$_0$ , $^3$F$_4$ and $^3$F$_3$ states at $E_{\rm x}^*$ = 21.0, 21.5, 25.7, and 26.7 MeV, respectively, from the analysis of the $\tau$+t elastic scattering.    
   
In this article, we report a kinematically complete $^3$H($\alpha$,$\tau$t)n experiment at the incident $\alpha$-particle energy of  67.2 MeV. This energy is sufficient to  excite $^6$Li levels over the $\tau$+t decay threshold of  $E_{\rm x}^*$ = 15.795 MeV, and thus the $\tau$+t  cluster states at about $E_x^*$ = 21 MeV can be investigated.  In our experiment, we  observed two resonances at $E_x^*$ = 21.30 and 21.90 MeV.

\section{Experimental set-up }

  The $^3$H($\alpha$,$^3$He\,$^3$H)n experiment was carried out by a target made of titanium backing saturated with tritium. The thickness of the titanium backing  was 2.6 mg/cm$^2$. By considering the saturation of the tritium atoms in the lattice of the titanium atoms,  we have that the ratio between the number of titanium atoms and the tritium one is approximately equal to 1. In these conditions the equivalent thickness of the tritium target is about 0.15  mg/cm$^2$. Therefore the total  thickness of the titanium backing and tritium target was about 2.75 mg/cm$^2$. The target was bombarded with $\alpha$-particles at the isochronous cyclotron accelerator U-240 of the Institute for Nuclear Research in Kiev. The beam energy was determined to be $E_\alpha$ = 67.2$\pm$0.4 MeV  by using a technique developed to measure time and energy  characteristics of the cyclotron beam \cite{Zerkin}.

We used a pair of $\Delta$E-E telescopes to detect t and $\tau$ in coincidence   from the $^3$H($\alpha$,$\tau$t)n reaction. The first telescope was placed at left side and consisted of $\Delta$E(400 $\mu$m thick totally depleted silicon surface barrier detector (SSD))  and E (NaI(Tl) with 20 mm$^{\phi}\times$ 20 mm$^{t}$) detectors, and the second one was placed at right side and consisted of $\Delta$E (90 $\mu$m SSD) and E (Si(Li) with 3\,mm$^{t}$) detectors.The first telescope could detect tritons as well as protons and deuterons, and the second telescope could detect $\tau$ and $\alpha$ particles together with protons, deuterons, and tritons of  low energies. We recorded the $\Delta$E and E energy information and the time-of-flight difference between two telescopes. The coincidence events within about 100 ns were recorded by a CM 1420 computer.

\begin{figure}[h]
\begin{center}
\vspace{-0.5cm}
\resizebox{0.9\textwidth}{!}{\includegraphics{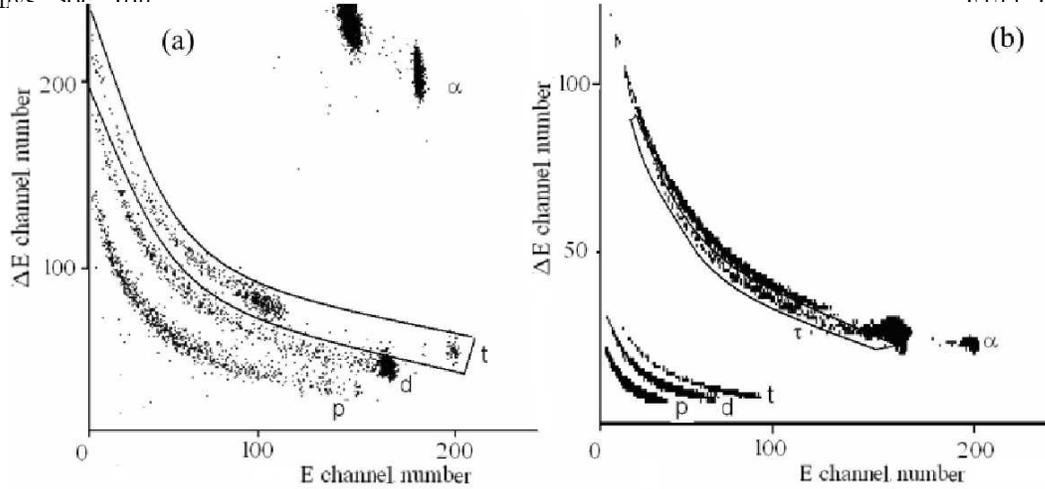}}
\end{center}
\vspace{-0.5cm}
\caption{Particle distributions in the ($\Delta E$, $E$)-plane related to the telescope placed at: a) $\theta_{\rm t}$ = 21$^\circ$ (the left side), b) $\theta_{\rm \tau}$ = 20$^\circ$( the right side), respectively, with respect to the beam axis direction. In the two squares are visible the various contributions of the detected particles.}
\label{fig2}
\vspace{-0.9cm}
\end{figure}

Figs. \ref{fig2}(a) and (b) show the correlation between $\Delta$E and  E for the first and second telescopes, respectively. It is seen that the t and $\tau$ particles from the $^3$H($\alpha$,$\tau$t)n reaction are clearly separated from other particles.

  The left telescope is optimized for the detection of single-charge particles such as p, d, and t. The $\alpha$-particles are also measured only for $E_\alpha >$ 29 MeV since those with lower energies are stopped in the $\Delta$E detector. In Fig. \ref{fig2} (a), high-energy $\alpha$-particles from the elastic scattering on Ti (backing) and $^3$H (target) form two peaks, but  $\alpha$-particles of low-energy could not be seen. On the contrary, the right telescope is optimized for the detection of double-charge particles such as $\tau$ and $\alpha$. In Fig. \ref{fig2} (b), the loci of p, d, and t are also seen but they are limited at low energies since those at high energies cannot be discriminated by the $\Delta$E counter.   
In the following, we express the total kinetic energy, $\Delta$E+E, as E for simplicity. 
   For the calibration of the scintillator used as E-detector placed at the $\theta_{\rm t}$  angle a special procedure based on the known Birks approach\cite{bircks51,bircks51p,bircks60,Gorpinich} was applied. Following relation (1) of Ref. \cite{bircks51} for the specific fluorescence $dS/dx$ of a material to an ionizing particle of energy $E$ with specific energy loss $dS/dx$, the scintillation response $S$ was obtained at large ionization energy losses ($KB\cdot dE/dx >> 1$) by expression 
  \begin{equation}
S = A \int_0^E \frac{dE}{1+KB \cdot dE/dx} \,\,\,\,\,\,.
\vspace{0.2cm}
 \label{eqbirk}
    \end{equation}

\begin{figure}[h]
\begin{center}
\resizebox{0.65\textwidth}{!}{\includegraphics{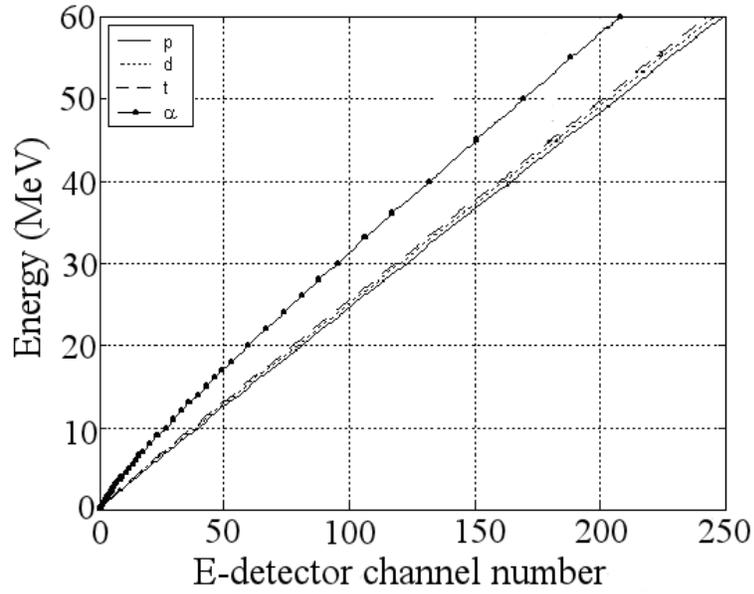}}
\end{center}
\caption{The energy dependence of the detected p, d, t, and $\alpha$ particles versus the channel number for the used NaI(Tl) scintillator.}
\label{fig3new}
\end{figure}

    For the given inorganic scintillator NaI(Tl), at energies of one- and two-charged  particles present in the spectrum of Fig. \ref{fig2} (a), we obtained the calibration  curves for the p, d, t, and $\alpha$ particles (see Fig. \ref{fig3new}).

\begin{figure}[h]
\begin{center}
\vspace{2.5cm}
\resizebox{1.1\textwidth}{!}{\includegraphics{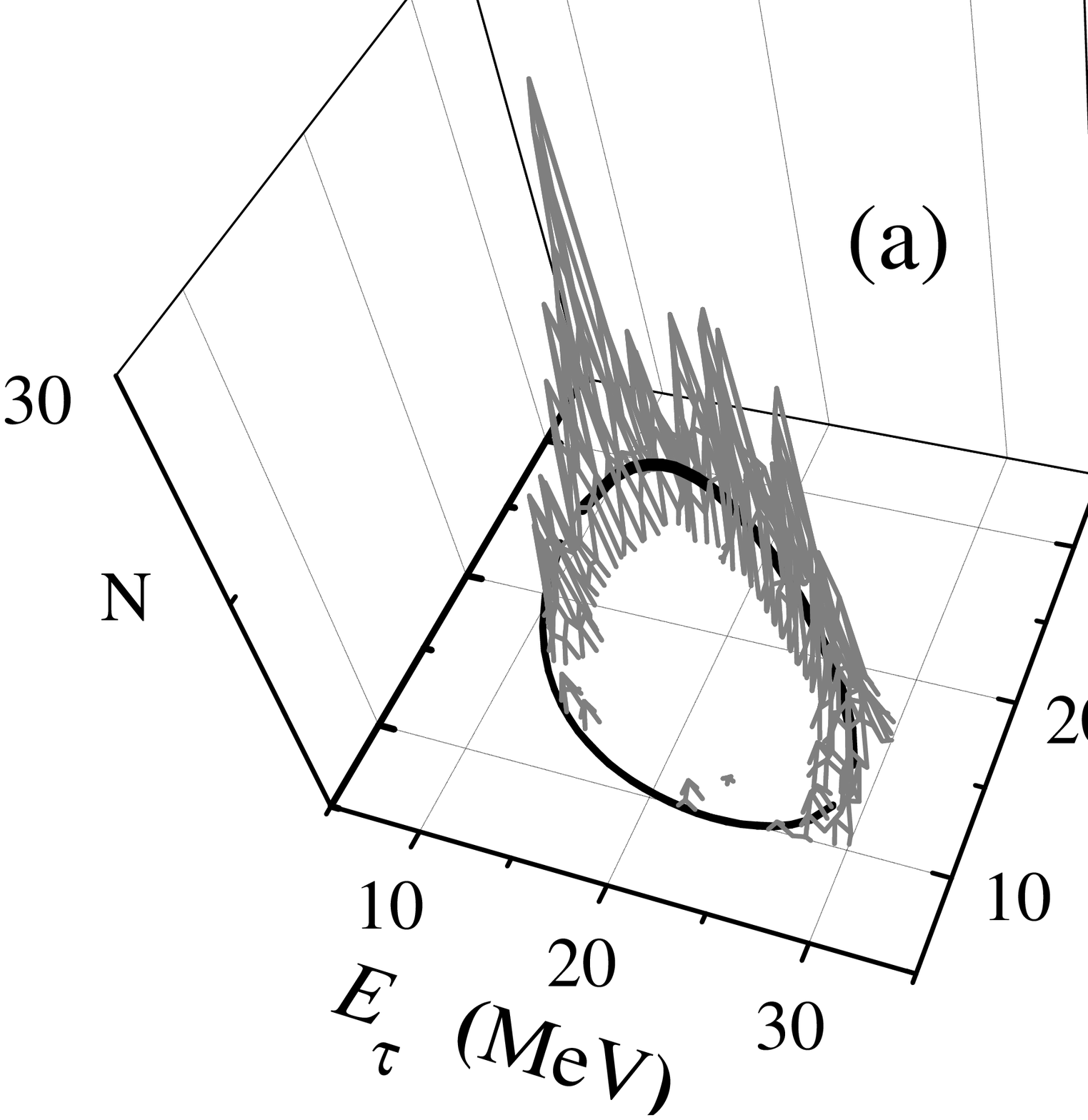}\includegraphics{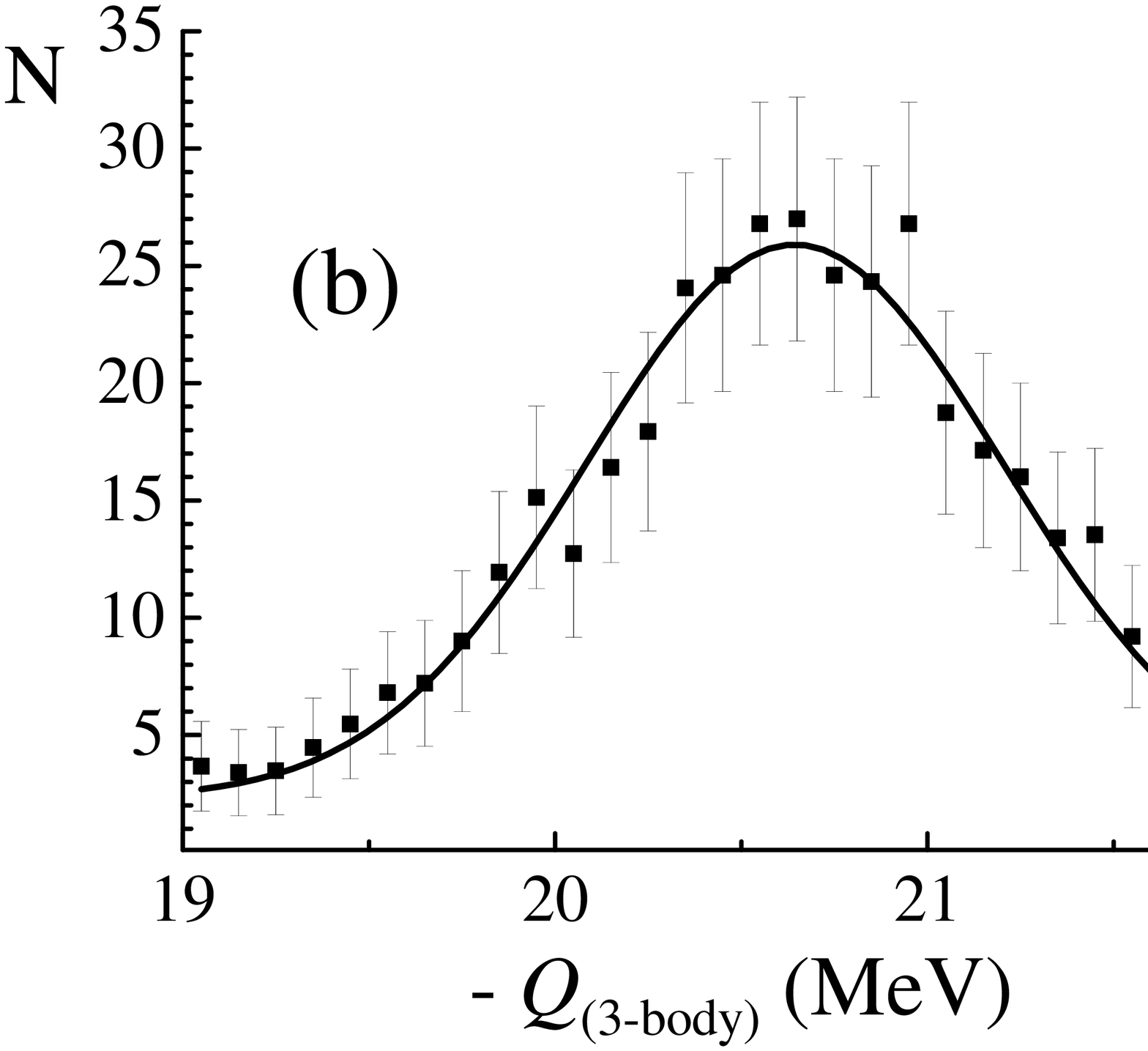}}

\end{center}
\vspace{-2.5cm}
\caption{ (a) Experimental bidimensional spectrum  of $\tau$t coincidences for the $^3$H($\alpha$,$\tau$t)n reaction at $\theta_{\rm\, \tau}$ = +20$^\circ$ and $\theta_{\rm t}$ = -21$^\circ$; the  solid line represents the kinematic curve calculated in the frame of a punctual geometry for the correspondent experimental  conditions. (b)  Experimental $Q$-value distribution for the three-body reaction obtained by the bidimensional spectrum analysis.}
\label{fig3}
\end{figure}

 Fig. \ref{fig3} (a) shows the $\tau$+t coincidence yield as functions of both the helium-3 energy $E_{\rm \tau}$ and the triton energy $E_{\rm t}$. The solid line represents the kinematical curve of the $^3$H($\alpha$,$\tau$t)n reaction in the present experimental condition.    

In order to check the reliability of the present experiment, the experimental Q-values for the 3-body reaction  were deduced by using the momentum and energy conservations \cite{Rae}: 
  
  \begin{eqnarray}
 \vec{P}_{\rm n}+\vec{P}_{\rm t}+\vec{P}_{\rm \tau}  & = & \vec{P}_{\rm \alpha}\label{eq1} \\
  E_{\rm n}+E_{\rm t}+E_{\rm \tau} + Q_{(\rm 3-body)} & = & E_{\rm \alpha} , \label{eq2}
\end{eqnarray}

  where $\vec{P}_{\rm n, t, \tau}$ and $E_{\rm n, t, \tau}$ are the momenta and energies of outgoing particles, respectively, while  $\vec{P}_{\rm \alpha_0}$ and $E_{\rm \alpha_0}$  are the momentum and energy of the incident $\alpha$-particle, respectively. Eq. (\ref{eq1}) can be used to calculate the momentum $P_{\rm n}$ and energy $E_{\rm n}$ of the undetected neutron, and then eq. (\ref{eq2}) allows to determine the  $Q_{(\rm 3-body)}$-value as $E_{\rm \alpha}$ - ($E_{\rm t}$ + $E_{\rm \tau}$ + $E_{\rm n})$. Taking into account the detector resolution, beam resolution, energy straggling in the target, effect of differential target thickness, kinematic changing from beam spot size and beam divergence, we obtain   the experimental $Q$-value peak of -20.61 MeV  for the $Q_{(\rm 3-body)}$ distribution (while the theoretical  $Q$-value  is  -20.58 MeV ) and the FWHM value of about 1.33 MeV (see Fig. \ref{fig3}(b)) with a standard deviation $\sigma$ of 0.56 MeV for the fit by a Gaussian function. This result shows the good agreement between measurements and calculation.

  \section{Analysis and results}

The obtained two-dimensional ($E_{\rm \tau}$,$E_{\rm t }$) spectra contain information on several reaction channels including the formation of $^6$Li$^*$. The formation of three particles, $\tau$+ t + n, in nuclear reactions is assumed as a sum of following contributions:

  \begin{eqnarray}
 ^3{\rm H} + \alpha & \rightarrow & \tau+ ^4{\rm H}^* \rightarrow  \tau + t + n \label{eq3} \\
              & \rightarrow  & n + ^6{\rm Li}^*     \rightarrow   n + \tau + t \label{eq4} \\
             & \rightarrow  & t + ^4{\rm He}^*       \rightarrow  t + \tau + n \label{eq5} \\
             & \rightarrow & n + {\rm quasifree\,\,} \tau+ t{\rm\, \,\,\, scattering}  \label{eq6}  \\
              & \rightarrow & \tau + t +n  \label{eq7}
\end{eqnarray}

where the processes (\ref{eq3}-\ref{eq5}) are the mechanisms of the subsequent decays, in which firstly unbound states of $^4$H$^*$, $^6$Li$^*$  and $^4$He$^*$ are formed and then these unbound states decay into the corresponding pairs of clusters. The process (\ref{eq6}) is the quasifree $\tau$+t scattering in which the $^3$He-particle comes from the virtual decay of $\alpha \rightarrow \,\tau$+n. The last process (\ref{eq7}) is the statistical three-body decay. The contribution from each process depends on the kinematics of the three-particle reaction. Therefore, the two-dimensional spectra obtained for different geometric conditions of the $\tau$-t coincidence were analyzed in order to find the best condition in which the $^6$Li$^*$ states with $\tau$+t cluster structure are significantly excited whereas the $^4$H and $^4$He resonance contributions are absent or are not overlapped with the ones of $^6{\rm Li}^*$. The triton and $\alpha$-particle detection angles are $\theta_t$ = -21$^\circ$ (left side) and $\theta_\tau$ = +20$^\circ$ (right side), respectively.

\begin{figure}[h]
\begin{center}
\resizebox{1.00\textwidth}{!}{\includegraphics{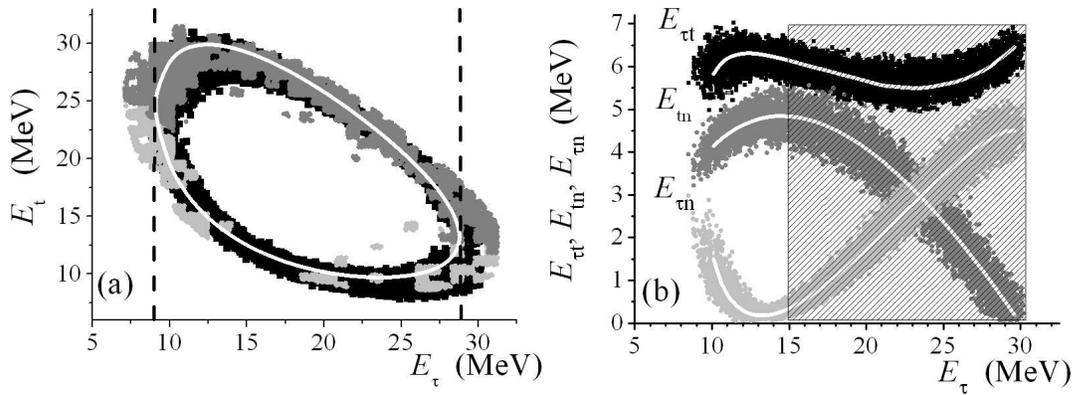}}
\end{center}
\vspace{-0.5cm}
\caption{(a) Experimental loci of the $\tau$t -coincidence events distributed in upper (intense gray region) and down (light gray region)  branches. The locus of the kinematic 3-body reaction calculation is marked with black background. The two vertical dotted lines, tangent to the two-dimensional three-body kinematic curve (dashed line), separates the upper branch from the lower one in the ($E_{\rm \tau}$, $E_{\rm t}$)-plane. (b)  Relative energies  $E_{\rm \tau t}$, $E_{\rm \tau n}$ and $E_{\rm t n}$ of the outgoing particle pairs versus $E_{\rm \tau }$ calculated in the frame of  a punctual geometry are marked as solid line. The same calculation with Monte Carlo is presented as colourful  arrays of dots. Shaded region corresponds to the 15-31 MeV   $E_{\rm \tau}$  energy range where the population and decay in the $\tau$+t clusters of excited $^6$Li levels occurs.}
\label{fig4}
\vspace{-0.5cm}
\end{figure}

  The typical  kinematic loci of the three-body reaction products, obtained by using the conservation laws expressed by equations (\ref{eq1}) and (\ref{eq2}), is presented in Fig. \ref{fig3}(a) (also including the yields of the coincidence events) and in Fig. \ref{fig4} (a) (the map of the two-dimensional three-body kinematic curve in the ( $E_{\rm \tau} $, $E_{\rm t } $)~--~plane). In general cases, for each $E_{\rm \tau}$ energy value correspond two real   $E_{\rm t}$ energy values, and the complete set of results covers the upper and lower branches of the ( $E_{\rm \tau} $, $E_{\rm t } $)~--~kinematic events. These upper and lower loci are separated by two vertical lines (see Fig. \ref{fig4} (a)) to which for each  $E_{\rm \tau} $ energy value corresponds two real and coincident solutions of the $E_{\rm t} $ energy value.

\begin{figure}[h]
\begin{center}
\vspace{3.8cm}
\resizebox{0.75\textwidth}{!}{\includegraphics{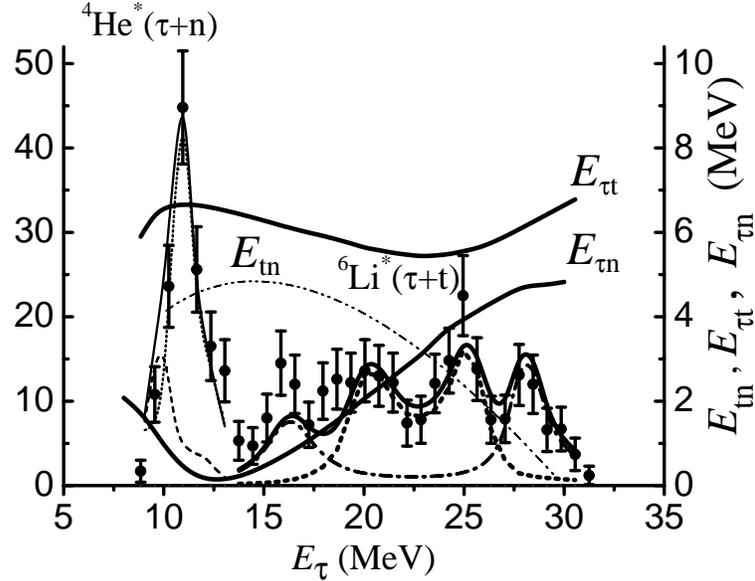}}
\end{center}
\vspace{-4.0cm}
\caption{Upper branch projection of the $\tau$-t coincidence event bidimensional spectrum onto the $E_{\rm \tau}$ axis obtained for $\theta_{\rm \tau}$ = 20$^\circ$  and $\theta_{\rm t}$ = 21$^\circ$ at incident energy of 67.2 MeV. The  $E_{\rm \tau t}$ function is the relative energy of the $\tau$-t system. The first peak (thin solid line) is due to the  decay of the  third (thin dashed line) and  second  (thin dotted line) excited $^4$He$^*$ levels. The successive four peaks are due to two different high-lying $^6$Li$^*$ levels (dash-dotted line is connected with the two contributions on the $E_{\rm \tau t}$ kinematic curve  corresponding to the $^6$Li$^*$  excited state at $E_{\rm x}$ = 21.90  MeV, while the dotted line is connected with the $^6$Li$^*$  excited state at $E_{\rm x}$ = 21.30  MeV). The thin dash-double dotted line is related to the $E_{\rm  t n}$ relative energy for the analysis of the $^4$H$^*$ excited state contributions.}
\label{fig5}
\vspace{-0.5cm}
\end{figure}

Moreover, among the various detection geometries, we have selected the one in which the relative kinetic energy $E_{\tau t}$ has a smooth slope as a function of $E_\tau$. This is a useful choice in order to determine the excitation energies of $^6$Li* with high accuracy. Unfortunately, due to the limited range of $E_{\tau t}$= 5.4--6.9 MeV (see Figs. \ref{fig4} (b) and \ref{fig5}), the explored $^6$Li* excitation energies are limited between $E_x^*$ = 21 and 23 MeV.Therefore, in the present experiment, the peak energies of $^6$Li* can be deduced with $\Delta$E $\leq$ 0.4 MeV whereas their width would be underestimated.

  In the present paper we  compare the experimental spectra with the simulation obtained by the Monte Carlo method. 
A three-body P(T,ab)c reaction is realized by using a set of random numbers suitable to obtain the a+b coincidence. In calculation we take into account for   the value of the beam energy and its dispersion, the thickness of the target, the energy loss  in target, the size of the spot beam on the target, the target-detector  distances and their energy resolutions.
  
  To analyze the experimental data coming from the $^3$H($\alpha$,$\tau$t)n reaction we should project the upper and lower loci of the kinematic curves onto the $E_{\rm \tau}$ and $E_{\rm t}$ energy axes. This procedure is made by recalculating the $\tau$t bidimensional spectra of the considered reaction by using the Monte Carlo method, as described in Ref.\cite{Povoroznyk}, and projecting the spectra onto the $E_{\rm \tau}$ or E$_{\rm t}$ axis with the consequently discontinuity due to the average of many values of a numerous set of random numbers.

   The selected ($E_{\rm \tau }$, $E_{\rm t }$) bidimensional spectra, obtained for the  incident energy $E_{\rm \alpha}$ of 67.2 MeV and detectors placed at $\theta_{\rm \tau}$ = +20$^\circ$ and $\theta_{\rm t}$ = - 21$^\circ$,  were divided in upper and lower branches (see Fig.\ref{fig4}(a)) by   using the above-mentioned method, and the upper branch of this locus is projected onto the $E_{\rm \tau}$ energy axis (see Fig. \ref{fig5}). Moreover, Figs. \ref{fig4}(b) and \ref{fig5} show the relative kinetic energies of the $\tau$-t, $\tau$-n and t-n pairs of particles versus the $E_{\rm \tau}$ energy.  In these  figures it is evident that the trend of the $E_{\rm \tau t}$ function  is almost constant with little fluctuation. 
   
The complete projection  onto the $E_{\rm \tau}$ axis of events populating the upper branch of the kinematic curve  is shown in Fig. \ref{fig5}. It is possible to observe the presence of five well resolved peaks due to the formation and decay of excited states of  $^4$He$^*$ and $^6$Li$^*$  nuclei. The error bars take into account the statistical error while the finite energy resolution of the $E_{\rm \tau}$ energy determination was 0.4 MeV. In order to avoid the overloading of the figure we do not include the horizontal error bars concerning the uncertainty of the  $E_{\rm \tau}$ energy.  
   In this figure, the first peak is due the contribution of $^4$He$^*$,  and the other four peaks are assumed to be due to the contributions of $^6$Li$^*$.

   In order to obtain the excitation energies and widths of $^6$Li$^*$, we performed a fitting with a Breight-Wigner formalism:

  \begin{equation}
   \frac{d^3 \sigma}{d\Omega_t d\Omega_\tau dE_\tau} \propto \rho(\Omega_t, \Omega_\tau, E_\tau) \sum^2_{j=1} C_j \frac{{(1/2\Gamma_j)}^2}{{(E_j-E_{\tau t})}^2 + {(1/2\Gamma_j)}^2}
   \label{eq8}
\end{equation}
   
where $C_j$, $E_j$ and $\Gamma_j$ are the amplitude, the peak energy  and width of the resonance $j$ ($j$=1,2), and  $\rho(\Omega_t, \Omega_\tau, E_\tau)$  is the phase space factor. The $\rho(\Omega_t, \Omega_\tau, E_\tau)$ and $E_{\tau t}(\Omega_t, \Omega_\tau, E_\tau)$ values were calculated by the Monte-Carlo method by taking into account for the present  experimental conditions. The dotted and dash-dotted lines represent the fits to the individual resonances while the solid line shows the sum of fitting. The peak energies are determined to be $ E_{\rm \tau t}$= 5.51 and 6.11 MeV. The the excitation energies of 
$^6$Li$^*$ can be obtained by  $E(^6{\rm Li}^*) = E_{\rm \tau t} + E_{\rm thr}$ with the threshold 
energy of $E_{\rm thr}$=15.79 MeV. Therefore we obtained two pairs of spectroscopic parameters:      
   $E^*_{\rm 1}$  = 21.30 $^{+ 0.3}_{-0.1}$ MeV with $\Gamma_{\rm 1}$ = 0.3 $^{+ 0.2}_{-0.1}$ MeV and $E_{\rm 2}$ = 21.90 $\pm$ 0.40 MeV with $\Gamma_{\rm 2}$ = 0.4 $\pm$ 0.2 MeV. The low error bars of $E^*_{\rm 1}$ and    $\Gamma_{\rm 1}$ determinations are smaller than the corresponding upper bars to cause of the much limited extension of the lower $E_{\tau t}$ relative energy values (5.4 MeV)  in comparison to the larger extension of the upper $E_{\tau t}$ values (6.9 MeV). Moreover, the measured $^6$Li$^*$ level widths are narrow and our results are, in any cases, lower than 0.6 MeV. This occurs because our experimental arrangement relatively to the investigated $^6$Li$^*$ levels allowed us to analyze only contributions coming from excitation energies around the peak values.
   However, the two above-mentioned  $^6$Li$^*$ levels exist and decay into $\tau$ and t clusters.
   Moreover, in the $E_{\rm \tau}$= 9-15 MeV  energy range of the projection of the upper branch,   where the $^4$He$^*$ decay in the $\tau$-n system appears, we extract the spectroscopic characteristics of the second and third excited levels  ($E_{\rm x}$ ($^4$He$^*$) with the respective $\Gamma$ width values).    
  In Fig. \ref{fig5} these excited $^4$He$^*$ states appear at the relative kinetic  energies of the $\tau$-n system of 0.45 and 1.25 MeV, respectively. The thin solid line  at $E_{\rm \tau}$ energies included in the above-mentioned energy range represents the whole contribution of the two excited $^4$He$^*$ states obtained by using the formalism (\ref{eq8}). These two contributions of the unbound $^4$He$^*$ states are represented by the dotted and dash-dotted lines. Therefore, the obtained spectroscopic parameters of the second and third $^4$He$^*$ excited states,  by considering the threshold of 20.60 MeV for the $^4$He$^*$ in the $\tau$+n system,  are $E_{\rm x}^*$ = 21.05 $\pm$ 0.10 MeV with width  $\Gamma$ = 0.45 $\pm$ 0.15 MeV, and $E_{\rm x}^*$ = 21.85 $\pm$ 0.10 MeV with width  $\Gamma$= 0.85 $\pm$ 0.25 MeV, respectively. These results are in agreement with the values reported by Tilley {\it et al.}\cite{Tilley92} in their compilation of the characteristics of nuclei with A = 4.

In order to explore if in the experimental spectrum of Fig. \ref{fig5} there are also present   contributions of $^4$H resonant states   we also report in the figure    the $E_{\rm t n}$ relative energies (thin dash-double dotted line)  of the t+n system. Considering that the energy threshold of the t+n system formation is placed at  21.108 MeV of the $^4$H excitation energy and that the line of the $E_{\rm t n}$ relative energy includes the interval between 0 and 4.9 MeV, only the high $^4$H  excited levels at 24.30, 24.61 and 27.13 MeV can populate the above-mentioned $E_{\rm t n}$ energy range at the 3.19, 3.50, and 5.27 MeV values, respectively. In the case of the 27.13 MeV excited state of $^4$H, the relative energy $E_{\rm t n}$ of 5.27 MeV overcomes the maximum value of 4.9 MeV reached for the $E_{\rm t n}$ relative energy and then only the spectrum tail of such an excited $^4$H is accessible in our explored $E_{\rm t n}$ energy range. The figure shows that the 3.19, 3.50, and 4.9 MeV of the $E_{\rm t n}$ relative energies correspond to the $E_{\rm \tau}$ energy values of 23.1, 22.0 and 14.8 MeV, respectively,  along the abscissa axis. As one can see, the experimental data of the total spectrum at these $E_{\rm \tau}$ energy values correspond to relative minimums. Therefore, we can state that in our experiment the possible $^4$H high resonant state do not give meaningful  contributions to the total recorded events which can belong to the reaction mechanism  (4), along the explored 3-body kinematic curve in  ($E_{\rm \tau}$ , in the $E_{\rm t}$) - plane.

     Moreover, Nakayama {\it et al.}\cite{Nakayama}, by the investigation of the $^7$Li($^3$He,$\alpha$) reaction forming $^6$Li,  found two resonant  $\tau$-t excited states at $E^*_{\rm x}$ = 18 $\pm$ 0.5 MeV with the width of 8 $\pm$ 1 MeV  and $E^*_{\rm x}$ = 22 $\pm$ 1 MeV with the width of 5 $\pm$ 1 MeV. These excited states were ascribed to the wide $^6$Li resonance at $E^*_{\rm x}$ of about 21 MeV\cite{Nakayama,Akimune}.  By the angular correlation analysis Nakayama {\it et al.}\cite{Nakayama} assigned both of the excited states at  $E^*_{\rm x}$ = 18 and 22 MeV to the spectroscopic characteristics $^1$P and $^3$P states, respectively. Therefore, the wide $^6$Li resonance is due to the overlapping of the two above-mentioned excited states. In the present work we did not produce angular correlation and however we assumed that our found excited states at 21.3 and 21.9 MeV are P states following the RGM calculation of  Thompson and Tang\cite{Thompson} and in agreement with the results and considerations of Nakayama {\it et al.}\cite{Nakayama}, Akimune {\it et al.}\cite{Akimune} and Yamagata  {\it et al.}\cite{Yamagata}. In addition, for a completeness of information we also report that such assignments are different from results obtained by the Mondragon and Hernandez paper\cite{Mondragon}, and  the CSRGM calculations of Ohkura {\it et al.}\cite{Ohkura}.

 In our experiment we find with enough precision two nearly $^6$Li levels characterized by two  excited states at $E_{\rm x}^*$ = 21.30 and 21.90 MeV. Actually, among the above-mentioned results concerning  the parameters of high-lying $^6$Li$^*$ levels there are some discrepancy. Therefore, other measurements, calculation and comparison with the known characteristics of the He, Li and Be isobar nuclei, with A = 6, have still to be made  in the next future.  Moreover, in the present investigation  we have also observed the decay of high $^4$He$^*$ excited states while the contributions coming from the $^4$H$^*$ excited states were absent.

         Our results on the two very near investigated $^6$Li levels are consistent with the values of levels present in the Aizenberg- Selove compilation\cite{Ajzenberg1984}.

            \section{Conclusion}
            
            \vspace{0.5cm}
            
            We performed a kinematically complete experiment for the $^3$H($\alpha$,$\tau$t)n reaction at the $E_\alpha$ incident energy of 67.2 MeV. The high-lying $^6$Li$^*$ states with the $\tau$-t cluster structure have been investigated in the  $E_x^*$=21-23 MeV excitation energy range. Two resonances at excitation energies of $E_{\rm 1}^*$ = 21.30  $^{+0.3}_{-0.1}$ MeV (with $\Gamma_{\rm 1}$ = 0.3 $^{+ 0.2}_{-0.1}$)  and $E_{\rm 2}^*$ = 21.90 $\pm$ 0.40 MeV (with $\Gamma_{\rm 2}$ = 0.4 $\pm$ 0.2) have been identified. The results of these peak energies are consistent with both the theoretical calculation  considering the cluster structure\cite{Thompson} of $^6$Li$^*$ and the result for the analysis of the $\tau$+t elastic scattering \cite{Ajzenberg1984}. However, the results  are inconsistent with the ones obtained by the recent investigation on the $^7$Li($^3$He,$\alpha$)$^6$Li reaction which reports a wider separation between the two corresponding excited levels  of about 4 MeV. Therefore, further experimental studies as well as more detailed theoretical analyses are needed to resolve the discrepancies.

\section*{Acknowledgment}
This work was supported by the National Academy of Science of Ukraine. The Italian authors would like to thank the Institute for Nuclear Research in Kiev for the fruitful cooperation.


\begin{thebibliography}{99} 
\bibitem{Ajzenberg1984} 
F. Ajzenberg-Selove.: Nucl. Phys. A {\bf 413} (1984) 1.

\bibitem{Tilley02} 
D.R. Tilley, C.M. Cheves, J.L. Godwin, G.M. Hale, H.M. Hofmann, J.H. Kelley, C.G. Sheu  
      and  H.R. Weller: Nucl. Phys. A {\bf 708} (2002) 3.

\bibitem{Nakayama} 
S. Nakayama, T. Yamagata, H. Akimune, M. Fujiwara, K. Fushimi, M.B. Greenfield, K. Hara,  
      K.J. Hara, H. Hashimoto, K. Ichihara, K. Kawase, H. Matsui, K. Nakanishi, M. Sakawa, M. 
      Tanaka and M. Yosoi: Phys. Rev. C {\bf 69} (2004) 041304.


\bibitem{Thompson} 
D.R. Thompson and Y.C. Tang: Phys. Rev. Lett. {\bf 19} (1967) 87.

\bibitem{Vlastou} 
R. Vlastou, J.B.A. England, O. Karban and S. Baird: Nucl. Phys. A {\bf 292} (1977) 29.
 
\bibitem{Zerkin} 
V. Zerkin, V. Konfederatenko, O.M. Povoroznyk, B.G.Struzho, and A.M.Shydlyk: Kiev 1991. Prepr. INR, Academy of Sciences of Ukraine, 91-11, in Ukrainian

\bibitem{bircks51} 
 J.B. Birks: Phys. Rev. {\bf 84} (1951) 364.

\bibitem{bircks51p} 
  J.B. Birks: Proc. Phys. Soc. A {\bf 64} (1951) 874.
  
  
\bibitem{bircks60} 
  J.B. Birks: IRE Tran. Nucl. Sci.  {\bf NS 7} (1960) 2. 

      
\bibitem{Gorpinich}
O.K. Gorpinich, O.M. Povoroznyk, O.O. Jachmenjov: Scientific papers of the Institute for nuclear research Kiev 2002. N° 1(7), p. 163-169, in Ukrainian

  \bibitem{Rae} 
 W.D.M.Rae, A.J.Cole, B.G.Harvey, R.G.Stokstad: Phys. Rev. C {\bf 30} (1984) 158. 

 \bibitem{Povoroznyk} 
 O.M. Povoroznyk. Nuclear Physics and Atomic Energy. Kiev 2007. N° 2(20), p. 131-139, in 
      Ukrainian

\bibitem{Tilley92} 
D.R. Tilley, H.R. Weller and G.M. Hale. Nucl: Phys. A {\bf 541} (1992) 157.

\bibitem{Akimune} 
 H. Akimune, T.Yamagata, S. Nakayama, Y. Arimoto, M. Fujiwara, K. Fushimi, K. Hara, M. 
        Ohta, A. Shiokawa, M.Tanaka, H. Utsunomiya, K.J. Hara, H.P. Yoshida and M. Yosoi: Phys.  
        Rev. C {\bf  67} (2003) 051302(R).  
        
        
\bibitem{Yamagata}
        T.Yamagata, H. Akimune, S. Nakayama et al.: Phys.  
        Rev. C {\bf  71} (2005) 064316.
        
       
\bibitem{Mondragon} 
 A. Mondragon and H. Hernandez: Phys. Rev.C {\bf 41} (1990) 1975. 

\bibitem{Ohkura} 
 H. Ohkura, T. Yamada and K. Ikeda: Prog. Theor. Phys. {\bf 94} (1995) 47.  


\end{thebibliography}
\end{document}